\documentclass{article} \usepackage{amssymb} \usepackage{amsmath}

\makeatletter
\@addtoreset{equation}{section}
\makeatother

\begin{document}

\renewcommand{\r}{\mathbf{r}}
\renewcommand{\k}{\mathbf{k}}
\newcommand{\Res}{\mathop{\text{Res}}}
\renewcommand{\atop}[2]{\genfrac{}{}{0pt}{}{#1}{#2}}
\newcommand{\FP}{\mathop{\text{FP}}}
\newcommand{\exc}{\text{exc}}
\newcommand{\cut}{\text{cut}}
\newcommand{\tlambda}{\tilde{\lambda}}

\title{General Considerations on the Finite-Size Corrections for
Coulomb Systems in the Debye--H\"{u}ckel Regime}

\author{Aldemar Torres and Gabriel T\'{e}llez \\ Universidad de Los
Andes\\ A.A. 4976\\ Bogot\'{a}, Colombia.}  

\date{}

\maketitle

\begin{abstract}
We study the statistical mechanics of classical Coulomb systems in a
low coupling regime (Debye--H\"uckel regime) in a confined geometry
with Dirichlet boundary conditions.  We use a method recently
developed by the authors which relates the grand partition function of
a Coulomb system in a confined geometry with a certain regularization
of the determinant of the Laplacian on that geometry with Dirichlet
boundary conditions. We study several examples of confining geometry
in two and three dimensions and semi-confined geometries where the
system is confined only in one or two directions of the space. We also
generalize the method to study systems confined in arbitrary
geometries with smooth boundary. We find a relation between the
expansion for small argument of the heat kernel of the Laplacian and
the large-size expansion of the grand potential of the Coulomb
system. This allow us to find the finite-size expansion of the grand
potential of the system in general. We recover known results for the
bulk grand potential (in two and three dimensions) and the surface
tension (for two dimensional systems). We find the surface tension for
three dimensional systems. For two dimensional systems our general
calculation of the finite-size expansion gives a proof of the
existence a universal logarithmic finite-size correction predicted
some time ago, at least in the low coupling regime. For three
dimensional systems we obtain a prediction for the perimeter
correction to the grand potential of a confined system.
\end{abstract}

\section{Introduction}

The study of confined classical Coulomb systems have attracted
attention for some time in particular because in some cases they
exhibit universal properties~\cite{Finite zise Gabriel, F-J-G
restricted dim, Finite zise jancov}. This universal behavior is
present in some correlations functions~\cite{janco corr-revisited} and
also on the thermodynamic quantities of the Coulomb systems. In
particular, the grand potential and thus the free energy, exhibit
finite-size corrections which depend on the confining
domain~\cite{Finite zise Gabriel}. Two particular cases have attracted
attention: fully confined systems and semi-confined systems. On this
respect we shall speak of semi-infinite or semi-confined systems to
refer to systems which are confined only in a certain spatial
direction (for example systems confined in a slab), to distinguish
them from the totally confined systems. In both cases, exactly
solvable models in two dimensions have allowed the explicit
calculation of the finite-size corrections in the free energy for a
given value of the coupling constant. For systems confined in a
metallic slab of width $W$, in $d$ dimensions, the free energy and the
grand potential per unit area (times the inverse reduced temperature)
exhibit an algebraic universal correction $C(d)/W^{d-1}$ with
\begin{equation}
  \label{eq:finite-size-corr-slab}
  C(d)=\frac{\Gamma(d/2)\zeta(d)}{2^d \pi^{d/2}}
\end{equation}
where $\Gamma(z)$ and $\zeta(z)$ are the Gamma function and the
Riemann zeta function respectively. This has been shown~\cite{Finite
zise Gabriel} to hold for any general Coulomb system provided that the
system is in a conducting phase and it has good screening
properties. It has also been checked in several solvable models. The
correction is universal in the sense that it does not depend on the
details of the microscopic constitution of the system.

For two-dimensional fully confined Coulomb systems there are also
universal finite-size corrections which are similar to those of two
dimensional critical systems. This lead us to another interesting
feature of conducting classical Coulomb systems, which is its manifest
similarity with critical systems~\cite{Finite zise Gabriel, Finite
zise jancov}. Although the particle and charge correlation functions
of the Coulomb system are short-ranged because of the screening, it
has been shown that the correlations of the electric field and the
electric potential are long ranged~\cite{janco corr-revisited,
Lebo-Martin}. In this sense they can be considered as critical systems
and they share properties of statistical models at criticality. For
example, in two dimensions conformal field theory, which has proved to
describe and classify correctly critical systems, predicts the
existence of universal corrections in the free energy for critical
systems due to their finite-size~\cite{cady finite size, cardy chap,
conformal}. Explicitly, for any two-dimensional statistical system in
its critical point confined in a domain of characteristic size $R$ with
smooth boundary, the free energy $F$ has a large-$R$ expansion of the
form~\cite{cady finite size, cardy chap}
\begin{equation}
  \label{eq:Cardy-finite-size}
  \beta F= A R^2 + B R -\frac{c\chi}{6}\ln R + \cdots
\end{equation}
where $\beta=1/(k_B T)$ with $T$ the absolute temperature and $k_B$
the Boltzmann constant. The first two terms $AR^2$ and $BR$ represent
respectively the bulk free energy and the ``surface'' (perimeter in
two dimensions) contribution to the free energy. In general, the
coefficients $A$ and $B$ are non-universal (they depend on the
microscopic detail of the model under consideration) but the
dimensionless coefficient of $\ln R$ is highly universal depending
only on the Euler characteristic of the manifold $\chi =2-2h-b$, where
$h$ is the number of handles and $b$ is the number of boundaries, and
on $c$ the central charge of the model. For Coulomb systems the
existence of a similar expansion, which reads
\begin{equation}
  \label{eq:Coulomb-finite-size}
  \beta F= A R^2 + B R +\frac{\chi}{6}\ln R + \cdots
\end{equation}
has been shown to hold in several exactly solvable models at a fixed
value of the coulombic coupling constant~\cite{Finite zise Gabriel,
Finite zise jancov, dosdesfera, Tellez-tcp-disque-neumann,
Tellez-Forrester-2dOCP-Gamma=4-6} and in some particular geometries
for any value of the coupling~\cite{Samaj-Janco-density-corr-TCP,
Janco-Trizac-sphere-correction-ocp, Janco-sphere-correction-tcp,
Samaj-asym-tcp-correction-sphere}.

In a previous paper~\cite{El nuestro} we considered two-dimensional
Coulomb systems in a low coupling regime, the Debye--H\"uckel
regime. We computed the grand potential for systems confined in two
simple geometries, the disk and the annulus with ideal conductor
boundaries, and confirmed the validity of the finite-size
expansion~(\ref{eq:Coulomb-finite-size}) in those cases. We showed
that the grand canonical partition function for a classical Coulomb
system in the Debye-H\"{u}ckel regime, confined with ideal grounded
conductor boundaries, can be expressed as an infinite product of
functions of the eigenvalues of the Laplace operator satisfying
Dirichlet boundary conditions. The explicit form of this spectrum and
the corresponding infinite products, depend on the shape of the
confining domain, and must be calculated for each particular
geometry. By a careful calculation of these infinite products we
obtained the explicit form of the grand potential for Coulomb systems
confined in a disk and in an annulus. When these systems are large we
computed the finite-size expansion of the grand potential and we found
the universal correction predicted by
Eq.~(\ref{eq:Coulomb-finite-size}). The first purpose of the present
paper is to apply this method to other particular cases of confining
geometry including semi-confined systems and also to systems in three
dimensions, for which conformal field theory predictions do not apply.

The second purpose this paper has to do with the fact that, from a
more general point of view, it is possible to define a spectral
function for the Laplacian, the heat kernel, that turns out to have an
asymptotic behavior for small argument which is independent of the
explicit form of the eigenvalues~\cite{Kac, curvature eigenv
laplacian}. Making use of these results, it is possible to show that
the spectrum of the Laplace operator calculated on a given manifold
endowed with a metric, contain geometrical information about the
manifold itself. In this paper we use those ideas to obtain the
large-size expansion of the grand potential for Coulomb systems
confined in arbitrary geometries.  Our results for the particular
cases agree with the predictions of this general formalism.

This paper is organized as follows. In
section~\ref{sec:previous-results} we summarize a few results of our
previous paper~\cite{El nuestro} concerning the calculation of the
grand potential for Coulomb systems in the Debye--H\"uckel regime in
given confining geometries. In particular, we briefly describe how the
grand potential can be obtained in terms of an infinite product of
functions of the eigenvalues of the Laplace operator.  In
sections~\ref{sec:examples} and~\ref{sec:general-case} we apply the
general method from Ref.~\cite{El nuestro} reviewed in
section~\ref{sec:previous-results}. In section~\ref{sec:examples} we
apply the method to some particular examples of confined and
semi-confined systems in two and three dimensions. In
section~\ref{sec:general-case} we consider the general case of fully
confined systems in an arbitrary geometry. We relate the grand
potential of the system to the zeta regularization of the determinant
of the Laplacian. By using the known results~\cite{Kac, curvature
eigenv laplacian} for the asymptotic expansion of the heat kernel we
find in general the finite-size expansion of the grand potential and,
for two-dimensional systems, we confirm the existence of the predicted
universal finite-size expansion. At the end of that section we present
an illustration of this latter method by considering the case of a
Coulomb system confined in a large square, and we recover a
finite-size correction predicted by conformal field
theory. Sections~\ref{sec:examples} and~\ref{sec:general-case} are
mostly independent and the reader not interested in the examples of
section~\ref{sec:examples} can proceed directly to the general
treatment exposed in section~\ref{sec:general-case}. In
section~\ref{sec:conclusion} we present a summary and gather some
conclusions.

\section{Summary of Previous Results}
\label{sec:previous-results}

Let us start by describing the model under consideration. Our system
is a multi-component Coulomb gas living in $d$ dimensions and composed
of $s$ species of charged particles $\alpha =1,...,s$ each of which
have $N_{\alpha }$ particles of charge $q_{\alpha }$. The system is
confined in a domain of volume $V$ with ideal conductor boundaries. We
shall describe describe this systems using classical
(i.e.~non-quantum) statistical mechanics in the grand canonical
ensemble. The average densities of the particles $n_{\alpha}$ are
therefore controlled by the fugacities $\zeta_{\alpha}$. We shall
impose the pseudo-neutrality condition
\begin{equation}
  \label{eq:pseudo-neutrality}
  \sum_{\alpha} q_{\alpha} \zeta_{\alpha}=0
\end{equation}
which implies that at the mean field level the system is neutral and
there is no potential difference between the system and the
boundaries. In the appendix B of Ref.~\cite{El nuestro} we explain
what happens in the more general case when the
condition~(\ref{eq:pseudo-neutrality}) is not satisfied.

The interaction potential between two unit charges located at $\r$ and
$\r'$ is given by the Coulomb potential $v(\r,\r')$ which is the
solution of Poisson equation
\begin{equation}
\label{eq:Poisson}
\Delta v(\r,\r')
=-s_{d}\delta (\mathbf{r}-\mathbf{r}^{\prime })  
\end{equation}
satisfying Dirichlet boundary conditions and where $s_{d}=2\pi
^{d/2}/\Gamma (d/2)$, that is in two dimensions $s_{2}=2\pi$ and for
three dimensional systems $s_{3}=4\pi$. For non-confined systems the
Coulomb potential reads
\begin{equation}
v^0(\mathbf{r},\mathbf{r}')
=
\begin{cases}
\displaystyle
\frac{1}{\left|
\mathbf{r}- \mathbf{r}'\right|}\,,
&\text{if\ } d=3\\
-\ln  \frac{\left|\mathbf{r}
-\mathbf{r}'\right|}{L}  \,,
& \text{if\ }d=2
\end{cases}
\end{equation}
where $L$ is an arbitrary length scale which fixes the zero of the
Coulomb potential in two dimensions. For the confined system under
consideration the explicit form of the Coulomb potential should be
modified in order to satisfy the Dirichlet boundary conditions.

As explained in Ref.~\cite{El nuestro} (see
also~\cite{Tellez-3Dslab}), the potential energy of the system can be
written as
\begin{eqnarray}
  \label{eq:hamiltonian}
H&=&\frac{1}{2}\sum_{\alpha ,\gamma }\sum\nolimits_{i,j}^{^{\prime
}}q_{\alpha }q_{\gamma }v(\mathbf{r}_{\alpha,i},
\mathbf{r}_{\gamma,j})+\frac{1}{2}\sum_{\alpha}\sum_{i}
q_{\alpha}^2 \left[
v(\mathbf{r}_{\alpha,i},\mathbf{r}_{\alpha,i})
-v^0(\mathbf{r}_{\alpha,i},\mathbf{r}_{\alpha,i})\right]
\nonumber\\
&=&
\frac{1}{2}\sum_{\alpha ,\gamma }\sum_{i,j}
q_{\alpha }q_{\gamma }v(\mathbf{r}_{\alpha,i},
\mathbf{r}_{\gamma,j})-\frac{1}{2}\sum_{\alpha}\sum_{i}
q_{\alpha}^2 
v^0(\mathbf{r}_{\alpha,i},\mathbf{r}_{\alpha,i})
\end{eqnarray}
In the first line the prime in the first summation means that the case
when $\alpha =\gamma $ and $i=j$ must be omitted. The first term is
the usual inter-particle energy between pairs. The second term is the
Coulomb energy of a particle and the polarization surface charge
density that the particle has induced in the boundaries of the system.

In Ref.~\cite{El nuestro} we performed the sine-Gordon
transformation~\cite{Sine gordon} on the grand canonical partition
function $\Xi$ of the system. Then we expanded the action around the
mean field solution to the quadratic order in the field. This is
valid in the Debye--H\"uckel regime. Then the remaining functional
integral can be performed easily since it is a Gaussian. The result is
a certain determinant involving the Laplacian which we put in the form
\begin{equation}
\Xi =\left(
\prod_{m}\left( 1-\frac{\kappa ^{2}}{\lambda _{m}}\right) 
\prod_{n}e^{\frac{
\kappa ^{2}}{\lambda _{n}^{0}}}\right) ^{-1/2}e^{\sum_{\alpha }V \zeta
_{\alpha }}.  \label{gp}
\end{equation}
where $\kappa ^{-1}=\left( \sum_{\alpha }s_{d}\zeta _{\alpha }\beta
q_{\alpha }^{2}\right) ^{-1/2}$ equals the Debye length in this
regime, $\lambda_m$ denotes the Laplacian eigenvalues satisfying the
Dirichlet boundary conditions and $\lambda _{n}^{0}=-\mathbf{K}^2$,
$\mathbf{K}\in\mathbb{R}^d$, refers to the (continuum) eigenvalues of
the Laplacian in the non-confined case. These come from the
``subtraction'' of the self-energy term $v^0(\r,\r)$ in
Eq.~(\ref{eq:hamiltonian}).

Each infinite product in (\ref{gp}) diverges separately. Indeed they
are ultraviolet divergent for large values of $|\lambda_{m}|$ and
$|\lambda_n^0|$. However, when they are putted together as in
(\ref{gp}), the divergences cancel each other (at least for the bulk
properties of the system). In three dimensions we can find immediately
a well defined expression for the grand potential from $\Omega
=-k_{B}T\ln \Xi$. In two dimensions, the situation is a bit more
involved since certain infrared divergence appear in the second
product and it must be regularized by introducing a lower cutoff. In
Ref.~\cite{El nuestro} we explained how to deal with this case and we
found the value of this cutoff explicitly in terms of the constant $L$
which fixes the zero of the Coulomb potential, which needs to be
supposed to be large. This cutoff was found to be given by $k_{\min
}=2e^{-C }/L$ where $C$ is the Euler constant. From Ref.~\cite{El
nuestro}, we recall that for a non-confined system Eq.~(\ref{gp})
gives for the bulk grand potential
\begin{eqnarray}
\frac{\beta \Omega_b }{V} &=&\frac{\kappa ^{2}}{4\pi }\left[ -\ln \frac{\kappa
L}{2}-C +\frac{1}{2}\right] -\sum_{\alpha }\zeta _{\alpha }
\qquad(2D)
\label{bulk2Dim} \\
\frac{\beta \Omega_b }{V} &=&-\frac{\kappa ^{3}}{12\pi }-\sum_{\alpha}
\zeta_{\alpha }
\qquad\qquad\qquad\qquad\quad\ 
(3D)
\label{bulk3Dim}
\end{eqnarray}
in two and three dimensions respectively. These expressions agree with
results by the usual formulation of the Debye-H\"{u}ckel
theory~\cite{Kennedy,Deutsch,Deutsch-Lavaud}.

\section{Solved Examples}
\label{sec:examples}

\subsection{Systems in Two Dimensions}

In Ref.~\cite{El nuestro} using Eq.~(\ref{gp}) we computed the grand
potential of a two-dimensional Coulomb system confined in a disk and
in an annulus, and we confirmed that its finite-size expansion is of
the form
\begin{equation}
  \beta \Omega = \beta \Omega_b + \beta \gamma + \frac{\chi}{6}\ln
  (\kappa R) + O(1)
\end{equation}
with $\Omega_b$ given by Eq.~(\ref{bulk2Dim}). The surface (perimeter)
tension is
\begin{equation}
  \label{eq:surface-tension-2D}
  \gamma = -k_B T \kappa/8\,.
\end{equation}
We notice the existence of the universal finite-size correction
$(\chi/6)\ln R$ with the Euler characteristic $\chi=1$ for the disk
and $\chi=0$ for the annulus.

In this section we will consider an additional example of confining
geometry in two dimensions.

\subsubsection{Space Between Two Infinite Lines: The Slab in Two
  Dimensions}
\label{sec:2Dslab}

The method outlined in section~\ref{sec:previous-results} can be
used to study semi-confined systems. In this subsection we
consider the case of such a system in two dimensions. The geometry
consists of two infinite parallel lines spaced by a distance $W$ and
the Coulomb systems is confined in between these two lines. We assume
Dirichlet boundary conditions for the electric potential. Let us
assume that the lines are in the direction of the $y$-axis and the
$x$-axis is perpendicular to the lines. If we write the Laplacian
eigenvalues as $\lambda=-k_x^2-k_y^2$, these take discrete values only
in the $k_{x}$ direction. The eigenfunctions can be written as $\Psi
(x,y)\varpropto e^{i(k_{y }y)}\sin(k_{x}x)$, satisfying the boundary
conditions $\Psi (0,y)=0$ and $\Psi (W,y)=0$, which imply
$k_{x}=n\pi/W$ with $n$ a positive, non-zero, integer. In the
direction of the $y$-axis there is no confinement therefore $k_{y}\in
\mathbb{R}$. Then, the eigenvalues of the Laplace operator are given
by $\lambda _{n,k_{y }}=-\left( n\pi/W\right)^{2} -k_{y}^{2}$, for
$n=1$, $2$, $\ldots\,$, and $k_{y }\in \mathbb{R}$. Introducing the
explicit form of the eigenvalues in (\ref{gp}) we have the grand
potential expressed as
\begin{equation}
\beta \Omega =\frac{1}{2}\frac{l}{(2\pi )}\int_{-\infty }^{\infty }\ln
\prod_{n=1}^{\infty }\left( 1+\frac{\kappa ^{2}}{\left( \frac{n\pi }{W}
\right) ^{2}+k_{y }^{2}}\right) dk_{y }+\frac{1}{2}\sum_{k}\frac{
\kappa ^{2}}{\lambda _{k}^{0}}-\sum_{\alpha }V \zeta _{\alpha }
\label{omega-slab2d}
\end{equation}
where $l$ is the dimension of the system in the $y$-direction. The
second term in~(\ref{omega-slab2d}) involve the spectrum for a
non-confined system, $\lambda^{(0)}=-\mathbf{K}^2$ with
$\mathbf{K}\in\mathbb{R}^2$. It can be written as
$\frac{1}{2}\sum_{k}\frac{\kappa ^{2}}{\lambda _{k}^{0}}=-\frac{ V
\kappa ^{2}}{4\pi }\int_{k_{\min }}^{K_{\max }}\frac{dK}{K}$, where $
V =lW$ is the ``volume'' (area) of the system between a portion of
length $l$ of the confining lines. The lower limit for this integral
is $ k_{\min }=2e^{-C }/L$ as mentioned in
section~\ref{sec:previous-results} and explained in Ref.~\cite{El
nuestro}. Also as explained earlier this integral is ultraviolet
divergent therefore we introduced an ultraviolet cutoff $K_{\max}$. In
the first term of Eq.~(\ref{omega-slab2d}) the infinite product
converges to a known expression~\cite{grad} giving
\begin{equation}
\label{eq:omega-2d-slab-simpl}
\beta \Omega =\frac{l}{2\pi }\int_{0 }^{k_{\max} }
\ln 
\frac{k_{y }\sinh
\left( W\sqrt{\kappa ^{2}+k_{y }^{2}}\right) }{\sqrt{\kappa
^{2}+k_{y }^{2}}\,\sinh (k_{y }W) }\,
dk_{y }
-\frac{ Wl
\kappa ^{2}}{4\pi }\int_{k_{\min }}^{K_{\max }}\frac{dK}{K}
-\sum_{\alpha }V \zeta _{\alpha }
\end{equation}
where we also introduced an ultraviolet cutoff for the first integral
$k_{\max}$. Both cutoffs $k_{\max}$ and $K_{\max}$ should be
proportional and their exact relation can be found by requiring that
in the limit $W\to\infty$ we recover the known bulk value of the grand
potential~(\ref{bulk2Dim}). Performing some of the integrals
in~(\ref{eq:omega-2d-slab-simpl}) we find that the grand potential per
unit length $\omega=\Omega/l$ is given by
\begin{eqnarray}
  \beta \omega
  &=&\frac{\kappa^2 W}{4\pi} \ln \frac{2k_{\max}}{K_{\max}}
  +\frac{\kappa^2 W}{4\pi}
  \left[
    \frac{1}{2} - \ln\frac{\kappa L}{2e^{-C }}
    \right]-W\sum_{\alpha}\zeta_{\alpha}\\
  && -\frac{\kappa}{4}
  +\frac{\pi}{24 W} 
  +\int_0^{\infty} \ln\left(1-e^{-2W\sqrt{\kappa^2+k_y^2}}\right)\,
  \frac{dk_y}{2\pi}
  \nonumber
\end{eqnarray}
where all terms that vanish when $k_{\max}\to\infty$ have been
omitted. Therefore to recover the known value~(\ref{bulk2Dim}) of the
bulk grand potential in the limit $W\to\infty$ the ultraviolet cutoffs
should be related by $K_{\max}=2k_{\max}$. Using these cutoffs we
finally find the grand potential and its finite-size expansion
\begin{subequations}
\begin{eqnarray}
\label{granp-slab2D}
\beta \omega&=&
\beta \omega_b+2\beta\gamma
+\frac{\pi}{24 W}
+\int_0^{\infty} \ln\left(1-e^{-2W\sqrt{\kappa^2+k_y^2}}\right)\,
\frac{dk_y}{2\pi}
\\
&=&\beta \omega_b+2\beta\gamma
+\frac{\pi}{24 W}
+O(e^{-2\kappa W})
\end{eqnarray}
\end{subequations}
where $\omega_b=\Omega_b/l$ with the bulk grand potential $\Omega_b$
given by Eq.~(\ref{bulk2Dim}). The surface tension $\gamma$ is given
by Eq.~(\ref{eq:surface-tension-2D}), which is the same surface
tension that we found in the case of the disk and the annulus in our
previous work \cite{El nuestro} as expected. Finally, we also found
the universal finite-size correction for the case of the slab in two
dimensions, which turns out to be $\pi/(24W)=\zeta (2)/(4\pi W)$ in
accordance with the general prediction from Ref.~\cite{Finite zise
Gabriel}: $\Gamma (d/2)\zeta (d)/(2^{d}\pi ^{d/2}W^{d-1})$ for $d=2$ as
expected.

\subsection{Systems in Three Dimensions}

In this section we consider some examples of three-dimensional Coulomb
systems first confined in a slab geometry then inside a ball and
inside a spherical thick shell.

\subsubsection{Space Between Two Infinite Planes: The Slab in Three
  Dimensions}
\label{sec:3Dslab}

We begin the study of particular examples of three-dimensional systems
by considering a system confined in the space between two infinite
parallel planes, separated by a distance $W$. Taking the
$x$-coordinate along the direction normal to the planes we find that
the eigenfunctions are $\Psi (\mathbf{r})\varpropto
e^{i(\mathbf{k}_{\perp }\cdot \mathbf{r}_{\perp })}\sin (k_{x} x),$
where $\mathbf{k}_{\perp }\cdot \mathbf{r}_{\perp }=yk_{y}+zk_{z},$
and satisfying the boundary conditions $\Psi (0,y,z)=0=\Psi
(W,y,z)$. Thus, the eigenvalues are given by $k_{x}=n\pi/W$, with
$n=1, 2, \ldots$, and $k_{y}\in \mathbb{R}$ and $k_{z}\in
\mathbb{R}$. Using (\ref {gp}) and the explicit form of the
eigenvalues we have
\begin{equation}
\beta \Omega =\frac{1}{2}\frac{A}{(2\pi )^{2}}\int \ln \prod_{n=1}^{\infty
}\left( 1+\frac{\kappa ^{2}}{\left( \frac{n\pi }{W}\right) ^{2}+\mathbf{k}
_{\perp }^{2}}\right) d\mathbf{k}_{\perp }+\frac{1}{2}\sum_{k}\frac{\kappa
^{2}}{\lambda _{k}^{0}}-\sum_{\alpha }V \zeta _{\alpha }
\label{omega-slab}
\end{equation}
where $A$ represents the area of the planes. The second term coming
from the subtraction of the self-energy is now
\begin{equation}
\frac{1}{2}\sum_{k}\frac{ \kappa ^{2}}{\lambda
_{k}^{0}}=-\frac{1}{2}\frac{\kappa ^{2}V }{(2\pi )^{3}}\int
d\tilde{\Omega} \int_{0}^{K_{\max
}}\frac{k^{2}dk}{k^{2}}=-\frac{\kappa^{2}V}{(2\pi)^2}
K_{\max }  
\end{equation}
where $V=AW$ is the volume of the system. As in the two-dimensional
example we introduced an ultraviolet cutoff $K_{\max}$.

Similarly to the two-dimensional slab, the infinite product in the
first term converges to~\cite{grad}
\begin{equation}
\prod_{n=1}^{\infty}
\left( 1+\frac{\kappa ^{2}}{\left( \frac{n\pi }{W}\right) ^{2}
+\mathbf{k}_{\perp}^2
}\right)=
\frac{ {k}_{\bot } }{\sqrt{\kappa
^{2}+\mathbf{k}_{\bot }^{2}}}\frac{\sinh \left( W\sqrt{\kappa
^{2}+\mathbf{k} _{\bot }^{2}}\right) }{\sinh (k_{\bot } W)}
\end{equation}
with $k_\perp=|\mathbf{k}_{\perp}|$. The remaining integral over
$\mathbf{k}_\perp$ is ultraviolet divergent and must be cutoff to a
maximum value $k_{\max}$ for $k_{\perp}$. In the limit
$k_{\max}\to\infty$ we have
\begin{eqnarray}
\frac{\beta \Omega }{A}&=&
\left[ 
\frac{\kappa^2}{8\pi}\left(k_{\max}-\frac{2K_{\max}}{\pi}\right)
-\frac{\kappa ^{3}}{12\pi }
-\sum_{\alpha }\zeta _{\alpha }
\right] W\\
&&+\frac{\kappa^2}{8\pi }\left[ \ln \frac{\kappa }{k_{\max }}
-\frac{1}{2}\right] \\
&&+
\frac{\zeta (3)}{16\pi W^{2}}
+\frac{1}{4\pi}\int_0^{\infty}
k\ln(1-e^{-2W\sqrt{k^2+\kappa^2}})\,dk
\end{eqnarray}
From this expression we see that the ultraviolet cutoffs should be
related by $k_{\max}=2K_{\max}/\pi$ in order to recover the known
value of the bulk grand potential~(\ref{bulk3Dim}) in the limit
$W\to\infty$. Then the grand potential is finally given by
\begin{subequations}
\begin{eqnarray}
\frac{\beta \Omega }{A}&=&
\hspace{-5mm}
\frac{\beta \Omega_b}{A}+
2\beta\gamma
+
\frac{\zeta (3)}{16\pi W^{2}}
+\frac{\kappa^2}{4\pi}\int_1^{\infty}
u\ln(1-e^{-2u\kappa W})\,du
\\
\label{eq:Omega-3Dslab-finite-size}
&\underset{W\to\infty}{=}&
 \frac{\beta \Omega_b}{A}+
2\beta\gamma
+
\frac{\zeta (3)}{16\pi W^{2}}
+O(e^{-2\kappa W})
\end{eqnarray}
\end{subequations}
with the bulk grand potential $\Omega_b$ given by Eq.~(\ref{bulk3Dim})
and the surface tension $\gamma$ given by
\begin{equation}
\beta\gamma=\frac{\kappa^2}{16\pi }\left[ \ln \frac{\kappa }{k_{\max }}
-\frac{1}{2}\right]
\underset{k_{\max}\to\infty}{=}
\frac{\kappa^2}{16\pi }\ln \frac{\kappa }{k_{\max }}
\end{equation}
Note that when we take the limit $k_{\max }\rightarrow \infty $ the
surface tension diverges with the cutoff as $-[\kappa^2/(16\pi)]\ln
k_{\max}$. This divergence in the surface tension can be understood if
we note that the particles tend to move to the frontiers because of
the ideal conductor character of the boundaries. This is easy to see
from a physical argument: the ideal conductor boundaries condition is
equivalent to introduce an image charge of opposite sign at the other
side of the boundary for each particle in the system. Particles near
the boundary ``feel'' an attraction to the boundary due to their
proximity with their corresponding images. Near a boundary, the
density of the species $\alpha$ at a distance $X$ from the boundary
will behave in this low coupling approximation as the linearized
Boltzmann factor of the particle-image interaction $1+\beta
q^2_{\alpha}/(4X)$. At large distances this interaction is screened
but at short distances it remains non-integrable. Since the surface
tension can be obtained as an integral of the density
profile~\cite{Samaj-Janco-TCP-metal}, this surface tension will be
infinite. Imposing a short-distance cutoff $D$ for the minimum
approach of the particles to the wall, will give a the surface tension
which diverges as $\ln D$. Our ultraviolet cutoff $k_{\max}$ is
proportional to $1/D$. For details see
Ref.~\cite{Tellez-3Dslab}. Notice that, on the other hand, for a
two-dimensional system the surface tension does not diverge with the
cutoff (see Ref.~\cite{El nuestro} and section~\ref{sec:2Dslab}). In
two dimensions the particle-image interaction is
$[q_{\alpha}^2/2]\,\ln(2X/L)$ and this expression is integrable at
short-distances. This explains why the surface tension is finite and
cutoff independent for two-dimensional systems although the particles
are strongly attracted to the boundaries contrary to the situation in
three dimensions where the surface tension diverges with the cutoff.

Returning to Eq.~(\ref{eq:Omega-3Dslab-finite-size}) we found a
finite-size correction depending on $W^{-2}$. This agree with the
universal finite-size correction for a slab in $d$-dimensions,
Eq.~(\ref{eq:finite-size-corr-slab}), for $d=3$ predicted in
Ref.~\cite{Finite zise Gabriel}.

\subsubsection{Coulomb System Inside a Ball}

We continue the study of finite-size Coulomb systems calculating the
grand potential for a three-dimensional Coulomb system confined inside
a spherical domain. The eigenvalue problem for the Laplace operator in
this case is easily solved.  The eigenfunctions are $\Psi (r,\theta
,\varphi )=\sqrt{\pi/(2r\lambda^{1/2})}I_{l+1/2}(\sqrt{\lambda
}r)Y_{l m}(\theta ,\varphi )$ where $I_{l+1/2}$ are the modified
Bessel functions of half integer order and $Y_{lm}$ are the spherical
harmonics. The eigenvalues $\lambda $ are determined from the
Dirichlet boundary condition $\Psi (R,\theta ,\varphi )=0$ where $R$
is the radius of the sphere.  Thus, the eigenvalues are the roots of
the equation $I_{l+1/2}(\sqrt{\lambda }R)=0.$ Let us call\ $\nu
_{l+1/2,n}$ the zeros of $I_{l+1/2}$. Then the eigenvalues are given
by $\lambda _{k}=\nu _{l+1/2,n}^2/R^2$ for $l=0,1,2\ldots$ and
$n=1,2,\ldots$. Also for each value of\, $l$ and $n$, the
corresponding eigenvalue is degenerated $2l+1$ times. Then, the
expression for the grand potential obtained from (\ref{gp}) takes the
form
\begin{equation}
\beta \Omega =\frac{1}{2}\ln \left[ \prod_{l=0}^{\infty }\left(
\prod_{n=1}^{\infty }\left( 1-\frac{z ^{2}}{\nu _{l+1/2,n}^{2}}\right)
\right) ^{2l+1}\right] +\frac{1}{2}\sum_{k}\frac{\kappa ^{2}}{\lambda
_{k}^{0}}-\sum_{\alpha }V \zeta_{\alpha }
\label{granpotencial-sphere}
\end{equation}
where the indexes $n$ and $l+1/2$ denote the root and the order of the
modified Bessel function $I_{l+1/2}$ respectively and $z =\kappa R$.
The infinite product over the index $n$ can be performed
exactly~\cite{grad}
\begin{equation}
\prod_{n=1}^{\infty }\left( 1-\frac{z ^{2}}{\nu _{l+1/2,n}^{2}}\right)
=\Gamma \left( l+ \frac{3}{2}\right) \left( \frac{2}{z }\right)
^{l+1/2}I_{l+1/2}(z ).
\end{equation}
The remaining summation over the index $l$ diverges and we must
regularize it by introducing an upper cutoff $N$ on $l$. 

The second term coming from the subtraction of the self energy is
\begin{equation}
\frac{1}{2}\sum_{k}\frac{ \kappa ^{2}}{\lambda
_{k}^{0}}=-\frac{1}{2}\frac{\kappa ^{2}V }{(2\pi )^{3}}\int
d\Omega \int_{0}^{K_{\max }}\frac{k^{2}dk}{k^{2}}=-\frac{\kappa
^{2}}{3\pi }R^{3}K_{\max }  
\end{equation}
where $V =\frac{4}{3}\pi R^{3}$ is the volume of the system. As in the
previous examples we introduced an ultraviolet cutoff $K_{\max}$ which
must be proportional to $N$ in order to cancel the divergences. The
exact relation between $K_{\max}$ and $N$ is found by the requirement
that the bulk value of the grand potential~(\ref{bulk3Dim}) is
recovered in the limit $R\to\infty$.

To find the finite-size expansion of the grand potential we make use
of the Debye uniform asymptotic expansion~\cite{Abramovitz} for the
Bessel functions, valid for large values of the argument,
\begin{equation}
\label{eq:Debye-ex-I_l}
\ln I_{\nu}(z )=-\frac{1}{2}\ln (2\pi )-\frac{1}{4}\ln \left( z
^{2}+\nu^{2}\right) +\eta (\nu,z )+\frac{3u-5u^{3}}{24\nu}+o\left( \frac{1}{
z ^{2}+\nu^{2}}\right)
\end{equation}
with
\begin{equation}
\label{eq:params-Debye-exp}
\eta (\nu,z)=\left( z ^{2}+\nu^{2}\right) ^{1/2}-\nu\sinh ^{-1}\left( 
\frac{\nu}{z }\right) ;\quad u=\frac{\nu}{\left( z
^{2}+\nu^{2}\right) ^{1/2}}
\end{equation}
together with the Euler-McLaurin summation formula to transform the
summation into an integral:
$\sum_{l=0}^{N}f(l)=\int_{0}^{N}f(l)dl+\frac{1}{2}\left[ f(0)+f(N)
\right] +\frac{1}{12}\left[ f^{\prime }(N)-f^{\prime }(0)\right]
+\cdots $, and the Stirling asymptotic expansion for the Gamma
function. Then in the limit $N\rightarrow \infty$ and $z \rightarrow
\infty$ we find
\begin{equation}
\beta
\Omega =\beta \Omega_{\text{bulk}}+\left( 1+2\ln \frac{R\kappa }{N}\right)
\frac{\kappa ^{2}R^{2} }{8}+\frac{\kappa R}{3}+O(R^{0})+o(N^{0})  
\end{equation}
where
\begin{equation}
\beta \Omega_{\text{bulk}}=\left[ -\frac{\kappa ^{3}}{12\pi }\allowbreak +\frac{
3\kappa ^{2}}{16\pi }\frac{N}{R}-\frac{\kappa ^{2}}{4\pi ^{2}}K_{\max
}-\sum_{\alpha }\zeta _{\alpha }\right] V  \label{bulk--sphere}
\end{equation}
This value of the bulk grand potential should be equal to the one
 given by Eq.~(\ref{bulk3Dim}) therefore $K_{\max }=(3\pi/ 4)N/R$.
 The ultraviolet cutoff $K_{\max}$ is indeed proportional to the
 cutoff $N$ and with this relation between the cutoffs the bulk
 divergences of the first and second terms in the
 r.h.s.~of~(\ref{granpotencial-sphere}) cancel each other. Reporting
 the result for $ K_{\max }$ in (\ref{bulk--sphere}) we find
\begin{equation}
  \beta \Omega =\left[ -\frac{\kappa ^{3}}{12\pi }\allowbreak
    -\sum_{\alpha }\zeta _{\alpha }\right] \frac{4\pi }{3}R^{3}+\left(
  \frac{\kappa^{2}}{32\pi } +\frac{\kappa^{2}}{16\pi}\ln \frac{\kappa
    R}{N}\right) 4\pi R^{2}+ \frac{\kappa }{3}R+o(R)
  \,.
  \label{gamx2}
\end{equation}
The first term in the r.h.s.~of Eq.~(\ref{gamx2}) is the bulk grand
potential in three dimensions. In the second term, the expression in
parenthesis gives the surface tension for the system and the third
term is a (non-universal) perimeter contribution. Notice again that
the surface tension diverges as $-[\kappa^2/(16\pi)]\ln K_{\max}$ with
the cutoff.

\subsubsection{Spherical Shell}

We consider now the case of a three-dimensional Coulomb system
confined inside a spherical shell.  Let $a$ and $b$ be the inner and
outer radii of the shell respectively. As in the previous example, the
eigenfunctions $\Psi (r,\theta ,\varphi )$ are separable in a radial
and an angular part. The eigenfunctions of the Laplacian for this
geometry are $\Psi (r,\theta ,\varphi )= \left[ A\sqrt{\frac{\pi
}{2r\lambda ^{1/2}}}I_{l+1/2}(\sqrt{\lambda }r)+B \sqrt{\frac{\pi
}{2r\lambda ^{1/2}}}K_{l+1/2}(\sqrt{\lambda }r)\right]
Y_{lm}(\theta ,\varphi ).$ The eigenvalues are determined by the
boundary conditions $\Psi (a,\theta ,\varphi )=\Psi (b,\theta ,\varphi
)=0,$ that is $ AI_{l+1/2}(\sqrt{\lambda }a)+B K_{l+1/2}(\sqrt{\lambda
}a)=0=AI_{l+1/2}( \sqrt{\lambda }b)+B K_{l+1/2}(\sqrt{\lambda }b)$.
These two equations can be considered as a linear system of equations
for the coefficients $A$ and $ B$. It has a non-trivial solution if
and only if
\begin{equation}
\label{eq:eigen-value-eq-shell}
I_{l+1/2}(\sqrt{\lambda }a)K_{l+1/2}( \sqrt{\lambda}b)-I_{l+1/2}
(\sqrt{\lambda }a)K_{l+1/2}(\sqrt{\lambda }b)=0
\,.
\end{equation}
The roots of this equation gives the eigenvalues. Let
$\vartheta_{l,n}$ be the $n$-th root of
$I_{l+1/2}(za)K_{l+1/2}(zb)-I_{l+1/2} (za)K_{l+1/2}(zb) =0$ for
$l=0,1,2\ldots\,$.  Then we have $\lambda _{k}=\vartheta _{l,n}^2$. Each
eigenvalue is $(2l+1)$-degenerated. Then the expression for the grand
potential takes the form
\begin{equation}
\beta \Omega =\frac{1}{2}\ln \left[ \prod_{l=0}^{\infty }
\prod_{n=1}^{\infty }\left(
1-\frac{\kappa^2}{\vartheta_{l,n}^{2}}\right)^{2l+1}\right]
+\frac{1}{2}\sum_{k}\frac{\kappa ^{2}}{\lambda _{k}^{0}}-\sum_{\alpha}
V \zeta_{\alpha }
\label{eq:granpotencial-shell0}
\end{equation}
The infinite product over the index $n$ can be performed explicitly by
using a method explained in
Refs.~\cite{Forrester-JSP,Tellez-tcp-disque-neumann,El nuestro}. Let us
consider the entire function
\begin{equation}
  f_l(z)=(2l+1)\frac{
    I_{l+1/2}(za)K_{l+1/2}(zb)-I_{l+1/2} (za)K_{l+1/2}(zb)
}{
    \left[ \left( \frac{a}{b}\right) ^{l+1/2}-\left(
    \frac{b}{a} \right) ^{l+1/2}\right]}
\end{equation}
which has the following properties: $f_{l}(z)=f_{l}(-z)$, $f_l(0)=1$,
$f_l'(0)=0$ and its zeros are $\vartheta_{l,n}$. Therefore it admits
an expansion as a Weierstrass infinite product
\begin{equation}
  f_{l}(z)=\prod_{n=1}^{\infty }\left(
  1-\frac{z^{2}}{\vartheta_{l,n}^{2}} \right)
  \,.
\end{equation}
Then the product we need to compute is simply $f_l(\kappa)$. The grand
potential then takes the form
\begin{equation}
\beta \Omega =\frac{1}{2}\sum_{l=0}^{N} (2l+1)
\ln f_l(\kappa)
-\frac{\kappa ^{2}}{3\pi }\left( b^{3}-a^{3}\right) K_{\max }
-\sum_{\alpha}V \zeta_{\alpha }
\label{eq:granpotencial-shell1}
\end{equation}
where we introduced the now familiar cutoffs $N$ for the sum and
$K_{\max}$ for the integral. As in all the previous examples these
cutoffs are proportional and their exact relation is found by
requiring that in the limit of a large system we recover the known
bulk value~(\ref{bulk3Dim}) of the grand potential.

For the calculation of the finite-size expansion of the grand
potential, notice first that the contribution of $K_{l}(\kappa
b)I_{l}(\kappa a)$ is of order $e^{-(b-a)\kappa }$, that is to say
exponentially smaller than the contribution of the term $I_{l}(\kappa
b)K_{l}(\kappa a)$ and as a consequence we can ignore it. To find the
value of the remaining summation over the index $l$, we use the
approximations of $K_{l+1/2}(\kappa a)$ and $I_{l+1/2}(\kappa b)$ for
large values of the argument, which can be obtained from the Debye
approximation for the modified Bessel functions~\cite{Abramovitz}.
Then we apply the Euler-McLaurin summation formula and take the limits
$N\to \infty$, $b\to\infty$, $a\to\infty$ and $b-a\rightarrow \infty$
with $a/b<1$ fixed. We find that the volume bulk part of the grand
potential is given by
\begin{equation}
\beta \Omega_{\text{bulk}}=\frac{N\kappa ^{3}}{4b\kappa }b^{3}-\frac{N\kappa ^{3}}{
4a\kappa }a^{3}-\left( \frac{\kappa }{9}+\frac{\kappa ^{2}k_{\max }}{3\pi }+
\frac{4\pi }{3}\sum_{\alpha }\zeta _{\alpha }\right) \left(
b^{3}-a^{3}\right)  \label{bulk-contr-shell}
\end{equation}
Equating this to the grand potential of the bulk system~(\ref{bulk3Dim})
we find 
\begin{equation}
K_{\max }=\frac{3\pi }{4} \frac{N}{b} 
\frac{1-(a^{2}/b^2)}{1-(a^{3}/b^3)}
\end{equation}
Using this relation we finally obtain 
\begin{equation}
\beta \Omega =\beta \Omega_{b}+\beta \Omega_{\text{surface}}+\frac{1}{3
}\left( b-a\right) \kappa +O(a^{0})+O(b^{0})
\end{equation}
where $\Omega_b$ is given by Eq.~(\ref{bulk3Dim}) and
\begin{equation}
\beta \Omega_{\text{surface}} =\frac{1}{8}\left( 2\ln \frac{a\kappa }{N}
-3\right) a^{2}\kappa ^{2}+\frac{1}{8}\left( 2\ln \frac{b\kappa }{N}
+1\right) b^{2}\kappa ^{2}
\end{equation}
Some additional comments are in order. The divergence in the surface
tension, familiar to us at this point, is also present in this
case. In the limit $K_{\max}\to\infty$ we have $\beta
\Omega_{\text{surface}}\rightarrow -4\pi \left(
b^{2}+a^{2}\right)[\kappa ^{2}/16\pi]\ln (K_{\max }/\kappa)$ which
allows us to define a surface tension $\gamma$ similar to the one of
the previous other three-dimensional cases
\begin{equation}
  \beta \gamma=-\frac{\kappa^2}{16\pi}\ln\frac{K_{\max}}{\kappa}
\end{equation}

We find again a perimeter contribution to the grand which is not
universal (it depends on the Debye length $\kappa^{-1}$). This time it
is given by $k_B T \kappa (b-a)/3$. This clearly suggests that this
perimeter correction is related to the curvature of the boundary and
probably to the curvature of the space itself. This is indeed the case
as we will show for any general geometry in the following section.

\section{Grand Potential for Arbitrary Confining Geometries}
\label{sec:general-case}

Up to now we have been capable to find the explicit form of the grand
potential for systems in specified confining geometries. Our
calculations of last section always involve the resolution of the
Laplacian eigenvalue problem for each specific geometry. From a more
general point of view it is possible to define functions of the
spectrum of the Laplacian that admit asymptotic expansions, that turn
out to have some properties that are independent of the explicit form
of the eigenvalues. Also in some cases these functions are related to
some invariants of the confining manifold, for example in two
dimensions to the Euler characteristic of the manifold. In this
section we make use of those ideas to find the finite-size expansion
of the grand potential in the case when the spectrum of the Laplacian
in the confining geometry is not known explicitly.

\subsection{Spectral Functions of the Laplacian}

In this section we review some spectral functions that will be useful
for our analysis. Let $\mathcal{M}_{g}$ be a Riemannian manifold
endowed with a certain metric $g$ and with boundary $\partial
\mathcal{M}_{g}$ and $\Delta $ the Laplace operator defined on
$\mathcal{M}_{g}.$ The spectrum of $\mathcal{M}_{g}$ is the set
$\left\{ 0\geqslant \lambda _{0}\geqslant \lambda _{1}\geqslant \cdots
\downarrow -\infty \right\} $ of eigenvalues of $\Delta$, that satisfy
$\Delta \Psi =\lambda \Psi ,$ where $ \Psi$ are the eigenfunctions of
$\Delta$. In order to determine the spectrum, these functions must
satisfy certain boundary conditions on $\partial \mathcal{M}_{g}$
which we impose to be of the Dirichlet type, that is $\Psi=0$ on
$\partial \mathcal{M}_{g}$. The first spectral function we are
interested in is the heat kernel defined as
\begin{equation}
\Theta (t)=\sum_{k=0}^{\infty }e^{t\lambda _{k}}
\,,
\end{equation}
which is convergent for $\Re e(t)>0$. It is known that the heat kernel
admits an asymptotic expansion for $t\to 0$ of the form
\begin{equation}
  \label{eq:HK-full-asympt-general}
  \Theta (t)\sim \sum_{n=0}^{\infty }c_{i_{n}}t^{_{i_{n}}}.  
\end{equation}
Here $\left\{ i_{n}\right\} $ is a certain increasing sequence of real
numbers and $i_{0}<0$. The exponent $i_{0}$ is particularly important
because $c_{i_{o}}t^{_{i_{o}}}$ is the divergent leading term in the
series. According to the famous Weyl estimate~\cite{Weyl} for the
Laplacian $i_0=-d/2$ where $d$ is the dimension of the manifold and
$(4\pi)^{d/2}c_{i_0}$ is the volume of the manifold. Following
Ref.~\cite{spectral functions} we define the order of the sequence of
the eigenvalues as $\mu =-i_{0}=d/2$. For a manifold $\mathcal{M}_g$
with a smooth boundary, some of the terms in the small-$t$ expansion
of the heat kernel have been found by Kac and others~\cite{Kac,
curvature eigenv laplacian}
\begin{equation}
(4\pi t)^{d/2}\Theta (t)=V
-\frac{\sqrt{4\pi t}}{4} B
+\frac{t}{6}(2C+D)+o(t^{3/2})  \label{expansion-HK}
\end{equation}
where 
\begin{eqnarray}
V &=&\text{the volume of }\mathcal{M}_{g} \\
B &=&\text{the surface area of }\partial \mathcal{M}_{g} \\
C &=& \text{ the curvatura integra }=\int_{\mathcal{M}
_{g}}K\\
D &=& \text{ the integrated mean curvature}=\int_{\partial 
\mathcal{M}_{g}}J
\end{eqnarray}
where $K$ is the scalar curvature at a point inside the domain
$\mathcal{M}_{g}$ and $J$ is the mean curvature at a point in the
boundary $\partial\mathcal{M}_{g}$. If we choose a metric $g$ in which
locally the first coordinate is perpendicular to the boundary and
outward pointing to it then the mean curvature $J$ can be computed
as~\cite{curvature eigenv laplacian}\footnote{Notice that we use here
the convention of outward pointing normal vectors to the
boundary. This is the opposite convention as the one in
Ref.~\cite{curvature eigenv laplacian}: our $J$ is minus the $J$ of
Ref.~\cite{curvature eigenv laplacian}.}
\begin{equation}
  \label{eq:mean-curvature}
  J=\partial_1[g^{11} \det g]
  \sqrt{g_{11}}/\det g
\end{equation}
Notice that in two dimensions the well-known Gauss--Bonnet
theorem \cite{diff geom} states that $ 2C+D=4\pi
\chi $ where $\chi $ is the Euler characteristic of the
manifold. Therefore in two dimensions the heat kernel expansion reads
\begin{equation}
\Theta (t)=\frac{V}{4\pi t}
-\frac{B}{8\sqrt{\pi t}} 
+\frac{\chi}{6} +o(t^{1/2})  \label{expansion-HK-2D}
\end{equation}


The second spectral function we are interested in is the Fredholm
determinant defined as $\prod_{k=0}^{\infty }\left( 1-\frac{ a
}{\lambda _{k}}\right)$ which is precisely the infinite product
involved in the calculation of the grand potential of the Coulomb
system from Eq.~(\ref{gp}). Unfortunately this infinite product only
converges for sequences of order $\mu<1$, and therefore it diverges
for the cases we are interested in, when $\mu=1$ (two dimensions) and
$\mu=3/2$ (three dimensions). For the cases $\mu>1$ a Weierstrass
canonical regularization of the Fredholm determinant
reads~\cite{spectral functions}
\begin{equation}
\digamma (a)=\prod_{k=0}^{\infty }\left(
1-\frac{a}{\lambda _{k}}
\right) \exp \left[ \frac{a}{\lambda_{k}}
+\frac{a^{2}}{2\lambda _{k}^{2}}+...+
\frac{ a^{[\mu ]}}{[\mu ]\lambda _{k}^{[\mu ]}}
\right]  \label{fred general}
\end{equation}
which is valid for $\mu >1$, where $[\mu ]$ is the integer part of
$\mu$.  The exponential term in (\ref{fred general}) is introduced in
order to make the infinite product convergent when the order of the
sequence is larger than one. We are interested in two- and
three-dimensional manifolds when $\mu$ equals $1$ or $3/2$
respectively. In both cases expression (\ref{fred general}) reduces to
\begin{equation}
  \label{fred-2d-3d}
  \digamma (a)= \prod_{k=0}^{\infty }\left(
  1-\frac{a}{\lambda_{k}}\right)
  e^{a/\lambda_{k}}.
\end{equation}
Although, strictly speaking the product~(\ref{fred general}) is only
defined for $\mu>1$ we will still use it in the two dimensional case
when $\mu=1$. We will see that using the
regularization~(\ref{fred-2d-3d}) for the ultraviolet divergence of
the Fredholm determinant in the case $\mu=1$ introduces some infrared
divergences. However, as we will show in detail later, these infrared
divergences can be dealt in a similar way as it was done for the two
dimensional examples of Ref.~\cite{El nuestro}.


We finally introduce the generalized zeta function defined as
\begin{subequations}
\begin{eqnarray}
  \label{eq:def-zeta}
  Z(s,a)&=&\sum_{k=0}^{\infty }\left( a-\lambda_{k}\right) ^{-s}
  \\
  \label{eq:zeta-from-heat-kernel}
  &=&\frac{1}{ \Gamma (s)}\int_{0}^{\infty }\Theta (t)e^{-at}t^{s-1}dt
  \,.
\end{eqnarray}
\end{subequations}
The first expression for the generalized zeta function defined as a
series is convergent for any $s$ such that $\Re e(s)>\mu$ and for any
$a$ such that $a\geqslant \lambda_{0}$. The second
expression~(\ref{eq:zeta-from-heat-kernel}) where the zeta function is
expressed as a Mellin transform of the heat kernel actually allows an
analytic continuation of $Z(s,a)$ for $\Re e(s)<\mu$ if the heat
kernel admits a full asymptotic expansion for $t\to0$ of the
form~(\ref{eq:HK-full-asympt-general}) as explained in
Ref.~\cite{spectral functions}. Note that $Z(s,0)=Z(s)$, where $
Z(s)=\sum_{k=0}^{\infty }\left( -\lambda_{k}\right) ^{-s}$ is the zeta
function of the sequence $\{\lambda_k\}$. The analytic continuation of
the zeta function has the following properties~\cite{spectral
functions} which we will need shortly. $Z(s)$ is meromorphic in the
whole complex $s$ plane and has poles at $s=-i_n$ with residue
\begin{equation}
  \label{eq:residue-zeta}
  \Res Z(-i_n)=\frac{c_{i_n}}{\Gamma(-i_n)}
  \,.
\end{equation}
In particular from Eq.~(\ref{expansion-HK}) we deduce that the first
pole is encountered at $s=-i_0=\mu=d/2$ and has residue
$V/[\Gamma(d/2)(4\pi)^{d/2}]$. Notice that this residue is independent
of the shape of the manifold: it only depends on its total volume $V$.
Also the negative or zero integers $s=-n$ are regular points of $Z(s)$
and we have~\cite{spectral functions}
\begin{equation}
  \label{eq:Z(-n)}
  Z(-n)=(-1)^n\,n!\,c_{-n}  
  \,.
\end{equation}
The generalized zeta function provides another regularization for the
Fredholm determinant of the Laplacian known as the zeta
regularization. Indeed differentiating~(\ref{eq:def-zeta}) under the
sum with respect to the variable $s$ and putting $s=0$ afterwards (a
procedure which is not legal since the expression~(\ref{eq:def-zeta})
is convergent only for $\Re e(s)>\mu$) formally yields
\begin{equation}
\left. \frac{\partial Z(s,0)}{\partial s}\right|_{s=0}
-\left. \frac{\partial Z(s,a)}{\partial s}\right|_{s=0}
=
\ln \prod_{k=0}^{\infty }\left( 1-\frac{a}{\lambda _{k}}\right)
\label{substration zetas}
\end{equation}
Strictly speaking this Eq.~(\ref{substration zetas}) is incorrect and
it should be only understood as a formal relation to justify the word
``determinant'' in the name zeta regularization of the determinant of
the Laplacian since the infinite product involved in the r.h.s.~is
divergent. Notice however the l.h.s. of Eq.~(\ref{substration zetas})
is well defined once the analytic continuation of $Z(s,a)$ is done
with the aid of relation~(\ref{eq:zeta-from-heat-kernel}).

The zeta regularization of the Laplacian determinant and the Fredholm
determinant~(\ref{fred-2d-3d}) are closely related. In
Ref.~\cite{spectral functions} a general relation between them is
found for any sequence of numbers of arbitrary order $\mu$. In our
particular case this relation reads
\begin{equation}
\label{eq:Fredholm-y-zeta-reg}
\left. \frac{\partial Z(s,0)}{\partial s}\right|_{s=0}
-\left. \frac{\partial Z(s,a)}{\partial s}\right|_{s=0}
=
\ln\left[\prod_{k=0}^{\infty }\left( 1-\frac{a}{\lambda _{k}}\right)
e^{a/\lambda_k}\right]
+a\FP[Z(1)]
\end{equation}
where $\FP[Z(1)]$ denotes the finite part of $Z$ at 1, defined as
usual by 
\begin{equation}
\FP[Z(s)]=
\begin{cases}
Z(s)& \text{if $s$ is not a pole of $Z$}\\
\displaystyle
\FP[Z(s)]=\lim_{\varepsilon\to 0}
\left[
Z(s+\varepsilon)
-\frac{\Res Z(s)}{\varepsilon}
\right]
&\text{if $s$ is a pole of $Z$}   
\end{cases}
\end{equation}

\subsection{The Connection with the Grand Potential of the Coulomb Systems}

As the reader probably noticed, the Fredholm determinant is quite
similar to the infinite product that appears in the expression for the
grand canonical partition function~(\ref{gp}) for Coulomb systems in
the Debye--H\"{u}ckel approximation. Our goal in this section is to
relate this kind of products with the geometrical information that
can be extracted from the asymptotic expansion of the heat kernel. 

\subsubsection{The bulk case}

First let us mention some points concerning the case of an unconfined
system. In this case the eigenvalues $\lambda_n=\lambda_n^{0}$ and the
expression~(\ref{gp}) for the grand canonical partition function
involves precisely the Fredholm determinant~(\ref{fred-2d-3d}) with
$a=\kappa^2$. The exponential terms $e^{\kappa^2/\lambda_n}$ that come
from the subtraction of the self-energy properly regularize the
infinite product $\prod_n[1-(\kappa^2/\lambda_n)]$. The final result
for the grand potential~(\ref{bulk2Dim}) and~(\ref{bulk3Dim}) is
finite and does not depend on any ultraviolet cutoff.

As a side note let us mention that if we were interested in the
formulation of Debye--H\"uckel theory for a system living in four or
more dimensions, the expression~(\ref{gp}) for the bulk grand
partition function would not be convergent and it would require the
introduction of an ultraviolet cutoff. This is because in dimension
$d\geqslant 4$ the index of the sequence of the Laplacian eigenvalues
would be $\mu=d/2\geqslant 2$. For this case the correct
regularization of the Fredholm determinant would require an additional
exponential term as shown in Eq.~(\ref{fred general}). Physically this
means that the bulk properties of a Coulomb system in dimension
greater or equal than four, in the Debye--H\"uckel regime, can only be
defined for a system of charged hard spheres or any other charged
particles with a short-range potential that cuts the singularity of
the Coulomb potential. The inverse of the radius of the particles is
the equivalent of the ultraviolet cutoff in our formulation. The bulk
thermodynamic properties would depend on the radius of the particles,
and diverge if one takes this radius to zero. This can be contrasted
with the two- and three-dimensional cases where one can build a
Debye--H\"uckel theory for which the bulk properties have a well
defined limit for point-like particles. 

As a complement on this remark let us remind the reader that for a
three-dimensional system the exact thermodynamic properties, beyond
the Debye--H\"uckel approximation, are not well defined for a system
of point-like particles due to the collapse of particles of opposite
sign. In two dimensions this collapse problem is less strong: if the
thermal energy of the particles is high enough a system of point
particles is well defined. On the other hand, the Debye--H\"uckel
approximation is less sensitive to this collapse problem: for two- and
three-dimensional systems the bulk properties are well defined for
point particles. However as we have seen in the examples, in the
three-dimensional case the surface properties are sensitive to the
collapse problem and a proper definition of these require the
introduction of a short-distance cutoff. For dimensions greater or
equal than four the collapse problem appears for the bulk properties
even in the Debye--H\"uckel approximation.

\subsubsection{Zeta regularized grand potential}

Now let us consider a confined Coulomb system in the manifold
$\mathcal{M}_g$. Let $R$ be the characteristic size of the
manifold. For instance one can define $R$ as $V^{1/d}$ where $V$ is
the volume of the manifold. We are interested in the large-$R$
expansion of the grand potential of the system, which can be obtained
from Eq.~(\ref{gp}). In this section we will study an intermediary
quantity $\Omega^{*}$ related to the grand potential which we define
by
\begin{equation}
  \label{eq:zeta-reg-grand-potential}
  \beta\Omega^{*}=\frac{1}{2}\left[Z'(0,0)-Z'(0,\kappa^2)\right]
  \,.
\end{equation}
For obvious reasons (remember Eqs.~(\ref{substration zetas})
and~(\ref{gp})) we will call this quantity the zeta regularized grand
potential. The prime in Eq.~(\ref{eq:zeta-reg-grand-potential})
indicates differentiation with respect to the first variable of the
zeta function ($s$).

As we will show below, the large-$R$ expansion of the zeta regularized
grand potential is related to the small-$t$ expansion of the heat
kernel~(\ref{expansion-HK}). To see this, let us consider a system
where all lengths have been rescaled by a factor $1/R$ : it is a
Coulomb system confined in a manifold of fixed volume equal to 1 but
with the same shape as the original system. Let $Z_{1}(s,a)$ be the
generalized zeta function for the spectrum of the Laplace operator for
such a manifold and $\Theta_{1}(t)$ its heat kernel. The subscript $1$
refers to a system confined in a volume 1. For the original system of
characteristic size $R$ we will eventually use the subscript $R$ in
the spectral functions $Z_{R}(s,a)$ and $\Theta_{R}(t)$.

The eigenvalues of the system of size $R$ are the same as those of the
system of size 1 multiplied by a factor $R^{-2}$. Then we have $\Theta
_{R}(t)=\Theta _{1}\left( R^{-2}t\right)$ and
$Z_{R}(s,a)=R^{2s}Z_{1}(s,aR^{2})$. From this we see that an expansion
of the heat kernel $\Theta_R$ for large-$R$ and fixed argument is the
same as an expansion of the heat kernel $\Theta_1$ for small argument.

We have
\begin{eqnarray}
\label{eq:Omega*1}
\beta\Omega^{*} &=&
\frac{1}{2}\left[\frac{ \partial }{\partial s}\left[
R^{2s}Z_{1}(s,0)\right]_{s=0}-\frac{\partial
}{\partial s}\left[ R^{2s}Z_{1}(s,\kappa^2 R^{2})\right]_{s=0}
  \right]
\\
&=& Z_{1}(0,0)\ln R- Z_{1}(0,\kappa^2 R^2)\ln R
-\frac{Z'_{1}(0,\kappa^2 R^2)}{2}
+\frac{Z_{1}'(0,0)}{2}
\nonumber
\end{eqnarray}
where the prime denotes differentiation with respect to the first
argument $s$. The last term is a constant, so we will eventually drop
it in the large-$R$ expansion.

From Eq.~(\ref{eq:zeta-from-heat-kernel}) we can
see that a small argument $t$ expansion of the heat kernel $\Theta(t)$
is equivalent to a large argument $a$ expansion of the zeta function
$Z(s,a)$. Then using the small argument expansion of the heat
kernel~(\ref{eq:HK-full-asympt-general}) into
Eq.~(\ref{eq:zeta-from-heat-kernel}) one can obtain in general the
large-$R$ expansion~\cite{spectral functions}
\begin{equation}
Z_{1}'(0,(\kappa R)^2)
\sim
\sum_{i_n\notin\mathbb{Z}^{-}\cup\{0\}}
c_{i_n} \Gamma(i_n) (\kappa R)^{-2i_n}
-\sum_{m=0}^{[\mu]}
c_{-m}\left[
\ln(\kappa R)^2-\sum_{r=1}^{m}r^{-1}\right]\frac{(-\kappa^2
  R^2)^{m}}{m!}
\,.
\end{equation}
In this equation the coefficients $c_{i_n}$ are those of the heat
kernel expansion~(\ref{eq:HK-full-asympt-general}) for a system of
size 1 and we use the convention that if $m$ is not any of the
exponents $i_n$ of Eq.~(\ref{eq:HK-full-asympt-general}) then
$c_m=0$. The first sum runs over all indexes $i_n$ that are not
negative or zero integers. Using this equation and~\cite{spectral
functions} $Z_{1}(0,a)=c_0-a c_{-1}$ into (\ref{eq:Omega*1}) yield the
large-$R$ expansion of the zeta regularized grand potential
\begin{multline}
  \label{eq:zeta-reg-gp*}
  \beta\Omega^{*}
  \underset{R\to\infty}{\sim}
  -\frac{1}{2}
  \sum_{i_n\notin\mathbb{Z}^-\cup\{0\}}
  c_{i_n} \Gamma(i_n) (\kappa R)^{-2i_n}
  -\kappa^2 R^2
  c_{-1} 
   \left[\ln\kappa-\frac{1}{2}\right]
  \\
  +c_0 \ln(\kappa R)
  +\frac{1}{2} Z'(0,0)
\end{multline}
Now we specialize this result for two and three dimensions. In two
dimensions from Eq.~(\ref{expansion-HK-2D}) we read
$c_{-1}=\tilde{V}/(4\pi)$, $c_{-1/2}=-\tilde{B}/(8\sqrt{\pi})$ and
$c_0=\chi/6$ where $\tilde{V}=V/R^2$ and $\tilde{B}=B/R$ denote the
volume (area) and perimeter of the manifold of characteristic size
1. Then we have
\begin{equation}
  \label{eq:zeta-reg-gp*-2D}
  \beta\Omega_{2D}^{*}= \frac{  \kappa^2 R^2}{4\pi}
  \left(\frac{1}{2}-\ln \kappa\right) \tilde{V}
  -  \frac{\kappa R}{8} \tilde{B}
  +\frac{\chi}{6}\ln(\kappa R)
  +O(1)
\,.
\end{equation}

In three dimensions from Eq.~(\ref{expansion-HK}) we have
$c_{-3/2}=\tilde{V}/(4\pi)^{3/2}$, $c_{-1}=-\tilde{B}/(16\pi)$ and
$c_{-1/2}=(2\tilde{C}+\tilde{D})/[6(4\pi)^{3/2}]$ with $\tilde{V}$ the
volume of the system of size 1, $\tilde{B}$ the area of the boundary
of system of size 1 and $\tilde{C}$ and $\tilde{D}$ the curvatura
integra and integrated mean curvature for the system of size
1. Replacing this into Eq.~(\ref{eq:zeta-reg-gp*}) we have
\begin{equation}
  \label{eq:zeta-reg-gp*-3D}
  \beta \Omega^{*}_{3D}= -\frac{\kappa^3 R^3}{12\pi} \tilde{V}
  +\frac{\kappa^2 R^2}{16\pi}\left[\ln
  \kappa-\frac{1}{2}\right]\tilde{B}
  +\frac{\kappa R }{48\pi}(2\tilde{C}+\tilde{D})
    +o(R)
\end{equation}

\subsubsection{Connection between the physical grand potential and the
  zeta regularized grand potential}

The excess grand potential $\Omega^{\exc}$ of the Coulomb system is
obtained from Eq~(\ref{gp}) as
\begin{equation}
  \label{eq:gp-exc}
  \beta\Omega^{\exc}=\frac{1}{2}\ln\left(
  \prod_{m}\left( 1-\frac{\kappa ^{2}}{\lambda _{m}}\right) 
  \prod_{n}e^{\frac{\kappa ^{2}}{\lambda _{n}^{0}}}\right)
\,.
\end{equation}
This expression involves a product very similar to the Fredholm
determinant~(\ref{fred-2d-3d}) but it only coincides with it for the
bulk case. In general for a confined system they are
different. However we can formally make appear the Fredholm
determinant $\digamma$ by writing
\begin{eqnarray}
  \beta\Omega^{\exc}&=&\frac{1}{2}\ln\left(
  \prod_{m}\left[\left( 1-\frac{\kappa ^{2}}{\lambda _{m}}\right) 
  e^{\frac{\kappa ^{2}}{\lambda _{m}}}\right]
  e^{\sum_{n}\frac{\kappa ^{2}}{\lambda _{n}^{0}}
    -\sum_{m}\frac{\kappa ^{2}}{\lambda _{m}}
  }\right)\nonumber\\
  &=&
  \frac{1}{2}\ln\digamma(\kappa^2)
  +\frac{\kappa^2}{2}
  \left[\sum_{m}\frac{1}{-\lambda_m}-\sum_{n}\frac{1}{-\lambda^{0}_n}
    \right]
  \,.
\end{eqnarray}
Then we make the connection with the zeta regularized grand potential
$\Omega^{*}$ defined in the previous section using
Eq.~(\ref{eq:Fredholm-y-zeta-reg}), so finally
\begin{equation}
  \label{eq:Omega*-Omega-gen}
  \beta\Omega^{\exc}=\beta\Omega^{*}
  -\frac{\kappa^2}{2}\FP[Z(1)]
  +\frac{\kappa^2}{2}
  \left[\sum_{m}\frac{1}{-\lambda_m}-\sum_{n}\frac{1}{-\lambda^{0}_n}
    \right]
  \,.
\end{equation}
The last two sums are divergent when taken separately. In principle
they should be cutoff in a similar way as in the examples. The proper
treatment of these sums should be done separately for each dimension
$d=2$ and $d=3$.

\subsubsection{Two-dimensional case}

In two dimensions the divergences of the sums involved in
Eq.~(\ref{eq:Omega*-Omega-gen}) can be dealt in an elegant way by
means of the zeta function. The zeta function has a pole at
$s=1$. Remembering Eq.~(\ref{eq:residue-zeta}) and the fact that the
residue of the zeta function $Z$ at $s=1$ for our confined system is
equal to the residue of $Z^{0}$ at $s=1$ for an unconfined system we
can identify the summations in Eq.~(\ref{eq:Omega*-Omega-gen}) with
\begin{equation}
  \sum_{m}\frac{1}{-\lambda_m}-\sum_{n}\frac{1}{-\lambda^{0}_n}
  =\lim_{s\to 1^{+}}\left[Z(s)-Z^{0}(s)\right]
\end{equation}
Then using the fact that both zeta functions $Z$ and $Z^{0}$ have the
same residue at $s=1$ we find that Eq.~(\ref{eq:Omega*-Omega-gen})
becomes
\begin{equation}
  \label{eq:Omega*-Omega-2D}
  \beta\Omega_{2D}^{\exc}=\beta\Omega^{*}_{2D}
  -\frac{\kappa^2}{2}\FP[Z^{0}(1)]
  \,.
\end{equation}
Now the zeta function for an unconfined system reads
\begin{equation}
  Z^{0}(s)=\sum_{n}\frac{1}{\left( -\lambda_{n}^{0}\right) ^{s} }
  =V\int_{\mathbb{R}^2}
  \frac{d^2\mathbf{k}}{(2\pi)^2}\frac{1}{\mathbf{k}^{2s} }
\end{equation}
However this zeta function cannot be properly defined: if $\Re
e(s)\leqslant 1$ the integral is ultraviolet divergent and if $\Re
e(s)>1$ it is infrared divergent. Depending on the sign of $\Re
e(s)-1$ this zeta function should be regularized with an ultraviolet
or infrared cutoff. For our present purposes we need $Z(s)$ defined
for $\Re e(s)>1$, then introducing the infrared cutoff
$k_{\min}=2e^{-C}/L$ as in the two dimensional examples of
section~\ref{sec:examples} and Ref.~\cite{El nuestro} we have
\begin{equation}
  Z^{0}(s)=\frac{V}{4\pi}\frac{k_{\min}^{2-2s}}{s-1}
\end{equation}
Its finite part at $s=1$ is
\begin{equation}
  \FP[Z^{0}(1)]=-\frac{V}{2\pi}\ln k_{\min}=
  -\frac{V}{2\pi}\ln \frac{2e^{-C}}{L}
  \,.
\end{equation}
Therefore
\begin{equation}
  \beta\Omega_{2D}^{\exc}=\beta\Omega^{*}_{2D}
  +\frac{\kappa^2 V}{4\pi}\ln \frac{2e^{-C}}{L}
  \,.
\end{equation}
Clearly this extra term only contributes to the bulk grand
potential. Finally reporting this into Eq.~(\ref{eq:zeta-reg-gp*-2D})
yields the finite-size expansion
\begin{equation}
  \label{eq:finite-size-exp-2D}
  \beta \Omega_{2D}^{\exc}= \frac{ \kappa^2 R^2}{4\pi}
  \left(\frac{1}{2}-C-\ln \frac{\kappa L}{2}\right) \tilde{V} -
  \frac{\kappa R}{8} \tilde{B} +\frac{\chi}{6}\ln(\kappa R) +O(1)\,.
\end{equation}

We have obtained the general finite-size expansion of the grand
potential for arbitrary confining geometry in two dimensions. We
recover from the first term of Eq.~(\ref{eq:finite-size-exp-2D}) the
bulk grand potential~(\ref{bulk2Dim}), from the second term the
surface tension already obtained in the examples $\gamma=k_B T
\kappa/8$ and finally the logarithmic finite-size correction
$(\chi/6)\ln R$. This constitutes a proof of the existence of this
universal finite-size correction for Coulomb systems, in the
Debye--H\"uckel regime, confined in an arbitrary geometry with
Dirichlet boundary conditions for the electric potential.

\subsubsection{Three-dimensional case}

For a three-dimensional system we must proceed differently to evaluate
the sums in Eq.~(\ref{eq:Omega*-Omega-gen}) as it was done in the
two-dimensional case. Here we cannot identify the sums with zeta
functions because in three dimensions the
definition~(\ref{eq:def-zeta}) of the zeta functions expressed as a
sum is only valid for $\Re e(s)>3/2$ and in
Eq.~(\ref{eq:Omega*-Omega-gen}) the sums correspond to $s=1$. In the
two-dimensional case we did not have this problem because the validity
of~(\ref{eq:def-zeta}) if for $\Re e(s)>1$ and since the residues of
both zeta functions are equal we could take the limit $s\to1^{+}$ of
the difference of zeta functions and obtain a finite result. Now, in
the three-dimensional case, we need the sums for a value $s=1<3/2$
which is far beyond the validity of Eq.~(\ref{eq:def-zeta}). Also the
corresponding analytic continuations of the zeta functions do not have
the same residue at $s=1$. Indeed the bulk zeta function $Z^0$ does
not have a pole at $s=1$ in the three-dimensional case but the zeta
function $Z$ for the confined system has a pole at $s=1$ with residue
given by Eq.~(\ref{eq:residue-zeta}) which is related to the
coefficient $c_{-1}$ corresponding to the surface contribution to the
grand potential. This suggest that the difference of the two sums in
Eq.~(\ref{eq:Omega*-Omega-gen}) will not be convergent for the
three-dimensional case as it was in the two-dimensional one and it
would give a surface contribution to the grand potential.

Let us introduce a truncated version of the zeta function evaluated at
$s=1$,
\begin{equation}
  Z_{\cut}(\tilde{\lambda})=\sum_{|\lambda_k|<\tilde{\lambda}}
  \frac{1}{-\lambda_k}
\end{equation}
and the corresponding one for the unconfined system
\begin{equation}
Z^{0}_{\cut}(\tilde{\lambda})
=\sum_{|\lambda_k|<\tilde{\lambda}}
\frac{1}{-\lambda_k}
=\frac{V}{(2\pi)^3}
\int_{|\k|^2<\tilde{\lambda}}
\frac{d^3\k}{\k^2}
=\frac{V}{2\pi^2}\tilde{\lambda}^{1/2}
\,.
\end{equation}
Here $\tilde{\lambda}>0$ is an
ultraviolet cutoff for the eigenvalues. The sums in
Eq.~(\ref{eq:Omega*-Omega-gen}) are
\begin{equation}
    \sum_{m}\frac{1}{-\lambda_m}-\sum_{n}\frac{1}{-\lambda^{0}_n}
    =\lim_{\tilde{\lambda}\to+\infty}\left[
	Z_{\cut}(\tilde{\lambda})-Z_{\cut}^{0}(\tilde{\lambda})
	\right]
      \,.
\end{equation}
Let us introduce the counting function $\mathcal{N}(\tilde{\lambda})$
which is equal to the number of eigenvalues $\lambda_k$ which are
$|\lambda_k|<\tilde{\lambda}$. The truncated zeta function is related
to the counting function by
\begin{equation}
  \label{eq:zeta-cut-N}
  Z_{\cut}(\tlambda)=\int_{0}^{\tlambda} 
  \frac{\mathcal{N}'(\lambda)}{\lambda}\,d\lambda
\end{equation}
with $\mathcal{N}'(\lambda)$ the derivative (in the sense of the
distributions) of $\mathcal{N}(\lambda)$. On the other hand the
derivative of the counting function and the heat kernel are related by
a Laplace transform
\begin{equation}
  \Theta(t)=\int_0^{\infty}
  e^{-t\lambda}\mathcal{N}'(\lambda)\,d\lambda
\end{equation}
Then, from the asymptotic expansion of the heat
kernel~(\ref{eq:HK-full-asympt-general}) for $t\to0$, we can obtain
the first terms of the asymptotic expansion of
$\mathcal{N}'(\tlambda)$ for $\tlambda\to+\infty$
\begin{equation}
  \mathcal{N}'(\tlambda)
  \underset{\tlambda\to\infty}{\sim}
   \frac{c_{-3/2}}{\Gamma(3/2)} \tlambda^{1/2}
   +c_{-1} + o(1)
   = \frac{V}{(2\pi)^2}\tlambda^{1/2}-\frac{B}{16\pi} + o(1)
\end{equation}
Reporting this into Eq.~(\ref{eq:zeta-cut-N}) we find the asymptotic
behavior of the truncated zeta function for $\tlambda\to\infty$
\begin{equation}
  Z_{\cut}(\tlambda)=\frac{V}{2\pi^2}\tlambda^{1/2}
  -\frac{B}{16\pi}\ln\tlambda + O(1)
\end{equation}
Then we have
\begin{equation}
  \label{eq:diff-zetas}
  Z_{\cut}(\tilde{\lambda})-Z_{\cut}^{0}(\tilde{\lambda})
  =-\frac{B}{16\pi}\ln\tlambda+O(1)=
  -\frac{B}{8\pi}\ln K_{\max}+{O}(1)
\end{equation}
where we introduced $K_{\max}=\tlambda^{1/2}$ to make the connection
with the examples of section~\ref{sec:examples}. We see that the
divergences are not completely canceled now as opposed to the
two-dimensional case. There remains an ultraviolet divergence which
contributes to the surface tension (proportional to $B$ the area of
the boundary of the manifold). This is the same kind of divergence
that we have found in the three-dimensional examples in
section~\ref{sec:examples} and it is due to the strong attraction of
the particles to their images as explained earlier. 

Finally collecting the results from Eqs.~(\ref{eq:zeta-reg-gp*-3D}),
(\ref{eq:Omega*-Omega-gen}), and~(\ref{eq:diff-zetas}) the finite-size
expansion of the grand potential reads
\begin{equation}
  \label{eq:Omega-exc-3D}
  \beta \Omega^{\exc}_{3D} = -\frac{\kappa^3 R^3}{12\pi} \tilde{V}
  +\frac{\kappa^2 R^2}{16\pi}\left[\ln
  \frac{\kappa}{K_{\max}}+{O}((K_{\max})^0)\right]\tilde{B}
  +\frac{\kappa R }{48\pi}(2\tilde{C}+\tilde{D}) +o(R)
  \,.
\end{equation}
From this very general calculation we recover the bulk grand
potential~(\ref{bulk3Dim}), the surface tension $\gamma = k_B T
(\kappa^2/16\pi) \ln [\kappa/K_{\max}]+{O}(1)$ which is
ultraviolet divergent. Also we find the perimeter correction to the
grand potential (the term proportional to $R$) which depends on the
curvatura integra $C$ and the integrated mean curvature $D$.

In the example of section~\ref{sec:examples} of the three-dimensional
Coulomb system confined in a ball we have found a perimeter correction
to $\beta\Omega$ equal to $\kappa R/3$. For a ball of radius one in
the flat space $\mathbb{R}^3$ the curvatura integra is $\tilde{C}=0$
and the mean curvature computed from Eq.~(\ref{eq:mean-curvature}) is
$\tilde{J}=4$ and the integrated mean curvature is
$\tilde{D}=16\pi$. Then the correction predicted by
Eq.~(\ref{eq:Omega-exc-3D}), $\kappa
R(2\tilde{C}+\tilde{D})/(48\pi)=\kappa R/3$, is in agreement with the
explicit result found in the example.  For the other example of the
Coulomb system confined in the thick spherical shell, the agreement of
our general result~(\ref{eq:Omega-exc-3D}) with the explicit
calculation of section~\ref{sec:examples} is also straightforward.

\subsection{System Confined in a Square Domain}

The above analysis is valid for a confining manifold with smooth
boundary. However it can easily be generalized for a manifold with a
boundary with corners. As an illustration let us consider the case of
a two-dimensional Coulomb system confined in an square domain of side
$R$ and subjected to Dirichlet boundary conditions for the electric
potential.

Conformal field theory predicts that in the case of a two-dimensional
critical system confined in a geometry with corners in the boundary,
it appears a contribution to the free energy (times the inverse
temperature) equal to $[\theta/(24\pi)](1-(\pi /\theta )^{2})\ln R$
for each corner with interior angle $\theta$ \cite{cardy chap}. In the
case of a square $ \theta =\pi /2$ and the contribution per corner
equals $\frac{\pi /2}{24\pi } (1-(2\pi /\pi )^{2})\ln R=-(1/16)\ln
R$. Then the total contribution of the four corners is $-(1/4)\ln
R$. If the similarity of conducting Coulomb systems with critical
systems holds then we should expect a finite-size correction to the
grand potential times $\beta$ equal to $(1/4)\ln R$ for Coulomb
systems.

The eigenvalues for this case can be found easily by separation of
variables expressing the Laplace operator in rectangular coordinates
and solving the eigenvalue equation. The spectrum for a system in a
square of side equal to 1 is given by $\lambda_{n,l}=-\pi^2\left(
n^{2}+l^{2}\right)$, $n=1,2,\ldots$ and $l=1,2,\ldots\,$. The heat
kernel is
\begin{equation}
\Theta _{1}(t)=\sum_{k=0}^{\infty }e^{t\lambda _{k}}=\left(
\sum_{n=1}^{\infty }e^{-\pi ^{2}n^{2}t}\right) ^{2}
\end{equation}
which can be expressed in terms of the Jacobi theta function
\begin{equation}
  \vartheta_3(u|\tau)=\sum_{n=-\infty}^{+\infty} e^{i\pi\tau n^2}
  e^{2nui}
\end{equation}
as
\begin{equation}
  \Theta_{1}(t)=\frac{1}{4}\left[\vartheta_3(0|i\pi t)-1\right]^2
\end{equation}
The heat kernel expansion for $t\to0$ can be found using Jacobi
imaginary transformation~\cite{Whittaker-Watson}
\begin{equation}
  \vartheta_3(u|\tau)=(-i\tau)^{-1/2}e^{u^2/(i\pi\tau)}
  \vartheta_3\left(\left.\frac{u}{\tau}\right|-\frac{1}{\tau}\right)
\,.
\end{equation}
This gives
\begin{equation}
  \Theta_1(t)=\frac{1}{4}
  \left[-1+
    (\pi t)^{-1/2}\left(1+2\sum_{n=1}^{\infty} e^{-n^2/t}\right)
    \right]^2
\end{equation}
The asymptotic expansion for $t\to 0$ is
\begin{equation}
  \Theta_{1}(t)=\frac{1}{4\pi t}-\frac{1}{2\sqrt{\pi t }}+\frac{1}{4}
  +{O}\left(\frac{e^{-1/t}}{t}\right)
\end{equation}
Comparing with expression~(\ref{expansion-HK-2D}) for a smooth
boundary we recognize the first two terms: the volume (area)
($\tilde{V}=1$) and the surface (perimeter) ($\tilde{B}=4$) terms
which are the same. The constant term on the other hand is now equal
to $1/4$. Applying the same argument developed above for the general
case to this heat kernel we see that this constant term is the one
that gives the coefficient of the logarithmic finite-size correction
for the grand potential. Then the finite-size expansion for this
geometry reads
\begin{equation}
  \label{eq:finite-size-exp-2D-square}
  \beta \Omega= \beta\Omega_b -
  \frac{\kappa B}{8} +\frac{1}{4}\ln(\kappa R) +o(\ln R)
\end{equation}
with $\Omega_b$ given by Eq.~(\ref{bulk2Dim}) with $V=R^2$ and the
perimeter of the square $B=4R$ with $R$ the length of a side.
The logarithmic finite-size correction is in agreement with the one
predicted by conformal field theory with the appropriate change of
sign.

\section{Summary and Perspectives}
\label{sec:conclusion}

We have illustrated with several examples how to apply the method of
Ref.~\cite{El nuestro} to find the grand potential of a Coulomb system
in the low coupling regime and confined by ideal conductor
boundaries. We considered several examples: in two dimensions the slab
and in three dimensions the slab, the ball and a thick spherical
shell. The method can easily be adapted to other geometries. In all
the examples we also computed the finite-size expansion of the grand
potential. For the slab geometries, in three and two dimensions, we
recover a universal algebraic finite-size correction predicted in
Ref.~\cite{Finite zise Gabriel}. For two-dimensional fully confined
systems the finite-size expansion exhibits a universal logarithmic
term similar to the one predicted for critical systems by conformal
field theory~\cite{El nuestro}.

We have also extended the method to confined systems of arbitrary
shape with a smooth boundary. For this general case we showed how the
heat kernel expansion for small argument for the considered geometry
is related to the large-size expansion of the grand potential of the
Coulomb system. From this, we recovered the expressions for the bulk
grand potential and the surface tension which agree with those found
in the specific examples. Regarding the finite-size corrections, in
the case of two dimensions we proved the existence of a universal
logarithmic finite-size correction for the grand potential times
$\beta$ equal to $(\chi/6)\ln R$ with $\chi$ the Euler characteristic
of the confining manifold. For three dimensional systems we also found
a general prediction for the perimeter correction to the grand
potential but it is not universal (it depends on the Debye length).

The general treatment for arbitrary confining geometry exposed here is
done for Dirichlet boundary conditions but can also be adapted for
other kind of boundary conditions. For example for ideal dielectric
boundary conditions, i.e.~Neumann boundary conditions for the electric
potential, the heat kernel expansion is very similar to
Eq.~(\ref{expansion-HK}) except that the surface term changes of
sign~\cite{curvature eigenv laplacian}. As a consequence the surface
tension of a Coulomb system confined by ideal dielectric boundaries
would be minus the surface tension of a Coulomb system confined by
ideal conductor boundaries in the low coupling regime. For a confining
geometry with smooth boundary the finite-size expansion will read as
in Eqs.~(\ref{eq:finite-size-exp-2D}) and~(\ref{eq:Omega-exc-3D}) for
two and three dimensions respectively except for a change of sign in
the surface tension term.

The method exposed in Ref.~\cite{El nuestro} and used here can be
extended to compute the density profiles and correlation functions. An
example of such application for the case of a slab geometry in three
dimensions can be found in Ref.~\cite{Tellez-3Dslab}.

\section*{Acknowledgments}

The authors would like to thank B.~Jancovici for a critical reading of
the manuscript and for his comments during various stages of the work
presented here. The authors acknowledge partial financial support from
Banco de la Rep\'ublica (Colombia),
ECOS-Nord/COLCIENCIAS-ICFES-ICETEX, and COLCIENCIAS project
1204-05-13625.

\end{document}